\documentclass[prl,longbibliography,twocolumn,superscriptaddress,amsmath,amssymb,verbatim]{revtex4-1}

\usepackage{graphicx}
\usepackage{lmodern}
\usepackage{epstopdf}
\usepackage[colorlinks=true,citecolor=blue,linkcolor=blue,urlcolor=blue,bookmarks=true]{hyperref}

\newcommand{\bro}{Zn-brochantite}
\begin{document}

\preprint{APS/123-QED}

\title{Field-Induced Instability of a Gapless Spin Liquid with a Spinon Fermi Surface}

\author{M. Gomil\v sek}
\affiliation{Jo\v{z}ef Stefan Institute, Jamova c.~39, SI-1000 Ljubljana, Slovenia}
\author{M. Klanj\v sek}
\affiliation{Jo\v{z}ef Stefan Institute, Jamova c.~39, SI-1000 Ljubljana, Slovenia}
\author{R. \v Zitko}
\affiliation{Jo\v{z}ef Stefan Institute, Jamova c.~39, SI-1000 Ljubljana, Slovenia}\author{M. Pregelj}
\affiliation{Jo\v{z}ef Stefan Institute, Jamova c.~39, SI-1000 Ljubljana, Slovenia}
\author{F. Bert}
\affiliation{Laboratoire de Physique des Solides, CNRS, Univ. Paris-Sud, Universit\'e Paris-Sacley, 91405 Orsay Cedex, France}
\author{P.~Mendels}
\affiliation{Laboratoire de Physique des Solides, CNRS, Univ. Paris-Sud, Universit\'e Paris-Sacley, 91405 Orsay Cedex, France}
\author{Y. Li}
\affiliation{Department of Physics, Renmin University of China, Beijing 100872, P. R.  China}
\author{Q. M. Zhang}
\affiliation{Department of Physics, Renmin University of China, Beijing 100872, P. R.  China}
\affiliation{Department of Physics and Astronomy, Shanghai Jiao Tong University, Shanghai 200240 and Collaborative Innovation Center of Advanced Microstructures, Nanjing 210093, P. R. China}
\author{A. Zorko}
\email{andrej.zorko@ijs.si}
\affiliation{Jo\v{z}ef Stefan Institute, Jamova c.~39, SI-1000 Ljubljana, Slovenia}

\date{\today}

\begin{abstract}
The ground state of the quantum kagome antiferromagnet Zn-brochantite, ZnCu$_3$(OH)$_6$SO$_4$, which is one of only a few known spin-liquid (SL) realizations in two or three dimensions, has been described as a gapless SL with a spinon Fermi surface.
Employing nuclear magnetic resonance in a broad magnetic-field range down to millikelvin temperatures, we show that in applied magnetic fields this enigmatic state is intrinsically unstable against a SL with a full or a partial gap.
A similar instability of the gapless Fermi-surface SL was previously encountered in an organic triangular-lattice antiferromagnet, suggesting a common destabilization mechanism that most likely arises from spinon pairing.
A salient property of this instability is that an infinitesimal field suffices to induce it, as predicted theoretically for some other types of gapless SL's.

\end{abstract}

\maketitle

Fermi-surface instability is one of the central concepts in condensed matter physics, responsible for diverse collective phenomena \cite{shankar1994renormalization}.
In metals, various symmetry-broken phases, e.g., the BCS superconducting, Peierls \cite{peierls1955quantum}, electronic nematic \cite{oganesyan2001quantum, harter2017parity} and itinerant antiferromagnetic \cite{berlijn2017itinerant} states, occur due to such instabilities.
An extension of the Fermi-liquid theory to Mott insulators leads to fermionic quantum spin liquids (SL's) \cite{zhou2016quantum}.
These are intriguing disordered, yet highly entangled states of matter that are characterized by effective low-energy charge-neutral fermionic quasiparticles known as spinons, which interact through emergent gauge fields \cite{lacroix2011introduction,kitaev2006anyons,balents2010spin, savary2016quantum,zhou2016quantum}.
In analogy to the Fermi-surface instabilities in metals, many SL's with different symmetries may be considered as being born out of the SL with a spinon Fermi surface (dubbed a spinon metal) \cite{galitski2007spin,barkeshli2013gapless}.
Since this parent state is gapless and thus exposed to perturbations and fluctuations, finding its rare realizations is challenging {\it per se}  \cite{watanabe2012novel,isono2014gapless,li_gapless_2014,shen2016evidence}.
Moreover, clarifying the nature of its experimentally observed instabilities  \cite{itou2010instability,gomilsek2016instabilities} by confronting numerous theoretical proposals \cite{galitski2007spin,barkeshli2013gapless,lee2007amperean,grover2010weak,metlitski2015cooper} with experimental facts represents an even greater challenge.
Clearly, identifying some common origin of such instabilities would be beneficial  for obtaining an in-depth understanding of the Fermi-surface instabilities in general. 

In this context, \bro, ZnCu$_3$(OH)$_6$SO$_4$, a representative of the paradigmatic two-dimensional geometrically frustrated quantum kagome antiferromagnet \cite{lacroix2011introduction}, is of particular interest.
Around 10~K, well below the average nearest-neighbor exchange interaction $J=65$~K \cite{li_gapless_2014},
it exhibits a spinon Fermi-surface SL state with Pauli-like kagome-lattice magnetic susceptibility $\chi_{\rm k}$ and with specific heat $c_{\rm p}$ increasing linearly with temperature \cite{li_gapless_2014}. 
Unexpectedly, this state progressively transforms when the temperature is lowered as $\chi_{\rm k}$ and $c_{\rm p}/T$ get gradually enhanced and saturate at 2--3-times larger values below $\sim$0.6~K \cite{li_gapless_2014, gomilsek2016muSR}.
This state remains stable down to the lowest experimentally accessible temperatures ($T/J \lesssim 3\cdot 10^{-4}$) \cite{gomilsek2016instabilities}.
The crossover within the spinon Fermi-surface SL state is likely associated with impurities originating from the 6--9\% Zn-Cu intersite disorder \cite{li_gapless_2014} and coupled to the kagome spins \cite{gomilsek2016muSR}, which could affect the spinon density of states at the Fermi level by pinning of spinon excitations \cite{gomilsek2016instabilities}.

Moreover, diverse field-induced instabilities were reported in a few two-dimensional SL candidates, like spin freezing in the archetype quantum kagome antiferromagnet herbertsmithite \cite{jeong2011field} and the organic triangular antiferromagnet $\kappa$-(BEDT-TTF)$_2$Cu$_2$(CN)$_3$ \cite{shimizu2006emergence}, and non-trivial symmetry breaking and/or topological ordering in another organic triangular antiferromagnet EtMe$_3$Sb[Pd(dmit)$_2$]$_2$ \cite{itou2010instability}.
In this light, studying the response of the SL with a spinon Fermi surface in \bro~to the magnetic field could help address the fundamental question about the stability of this gapless state against time-reversal symmetry breaking. 
Specifically, are gapless excitations of this state also intrinsically unstable against the applied field, as theoretically predicted for some other SL's, like the Dirac $U(1)$ SL on the kagome lattice \cite{ran2009spontaneous}, or the Kitaev SL on the honeycomb \cite{kitaev2006anyons} and the hyperhoneycomb lattices \cite{hermanns2015weyl}?
 
Here we provide the answer to this question by reporting a field-induced instability of the gapless spinon Fermi-surface ground state in \bro.
This was discovered by employing nuclear magnetic resonance (NMR), which has proven particularly well-suited for differentiating between impurity and intrinsic properties of this compound, as the $^2$D nuclei dominantly couple to the kagome spins \cite{gomilsek2016instabilities}.
By performing a series of $^2$D NMR spin-lattice relaxation ($1/T_1$) measurements in various applied fields down to millikelvin temperatures, we find that, surprisingly, the critical temperature $T_{\rm c}$ associated with this instability scales linearly with the applied magnetic field $B$, yielding $T_{\rm c}\rightarrow 0$ at $B\rightarrow 0$. 
We propose that the field-induced SL state originates from spinon pairing and has a full or a partial excitation gap.
The instability is reminiscent of that found in the triangular-lattice gapless spinon Fermi-surface SL candidate EtMe$_3$Sb[Pd(dmit)$_2$]$_2$ and thus appears to be a common feature of such SL's. 
\begin{figure}[t]
\includegraphics[trim = 0mm 0mm 0mm 0mm, clip, width=1\linewidth]{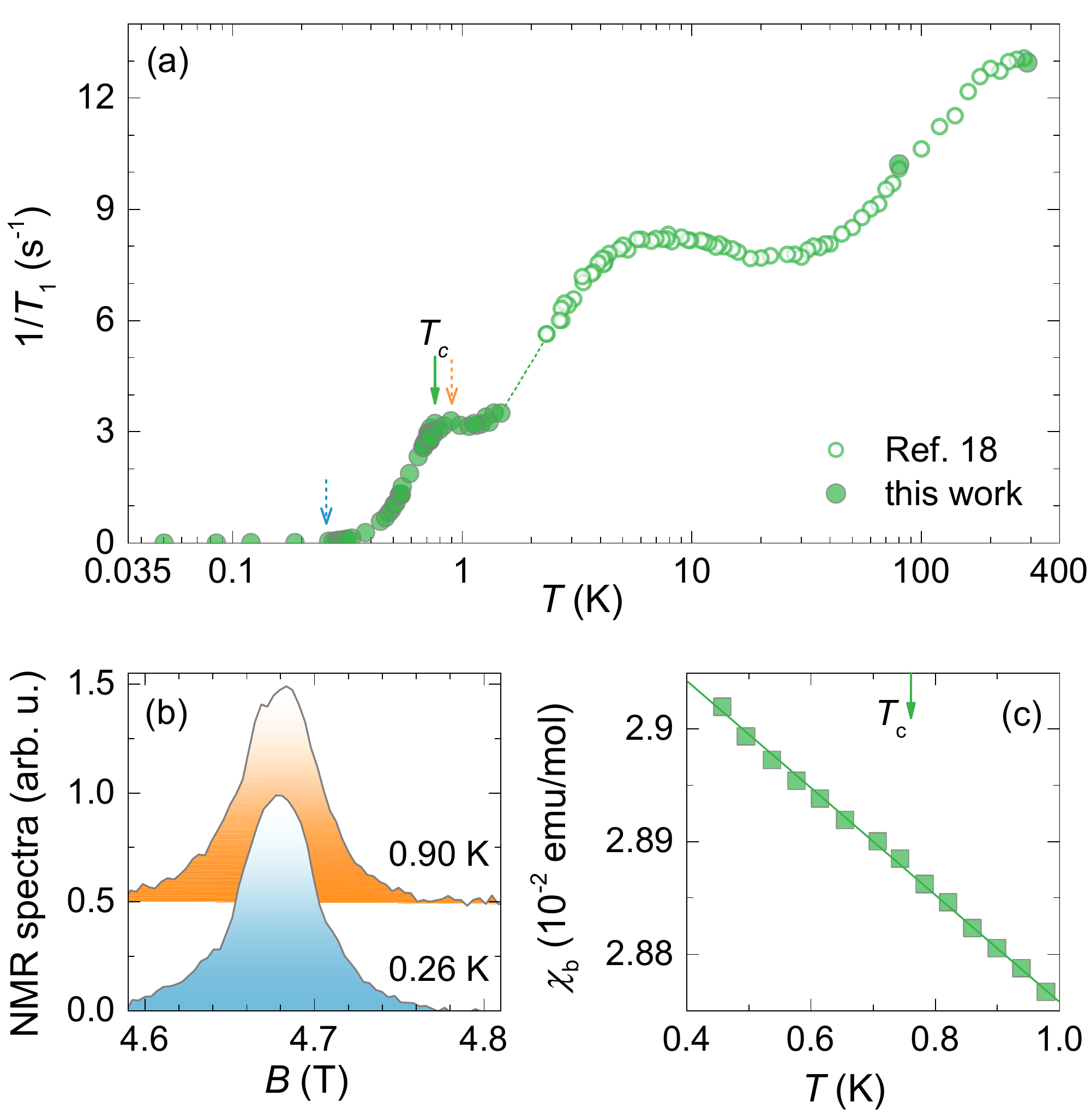}
\caption{(a) The temperature dependence of the $^2$D NMR spin-lattice relaxation rate $1/T_1$ in the applied field of 4.69~T.
The solid arrow indicates the transition temperature $T_{\rm c}=0.76$~K, while the dashed arrows indicate the temperatures where the $^2$D NMR spectra from panel (b) were recorded.
These spectra are normalized and shifted vertically for clarity.
(c) The temperature dependence of the bulk susceptibility $\chi_{\rm b}$ in 4.69~T measured by a SQUID magnetometer. 
Note that the decrease of $\chi_{\rm b}$ with temperature is very small. 
$T_{\rm c}$ is indicated by the arrow, while the solid line is a guide to the eye.
}
\label{fig1}
\end{figure}

\begin{figure}[b]
\includegraphics[trim = 0mm 0mm 0mm 0mm, clip, width=1\linewidth]{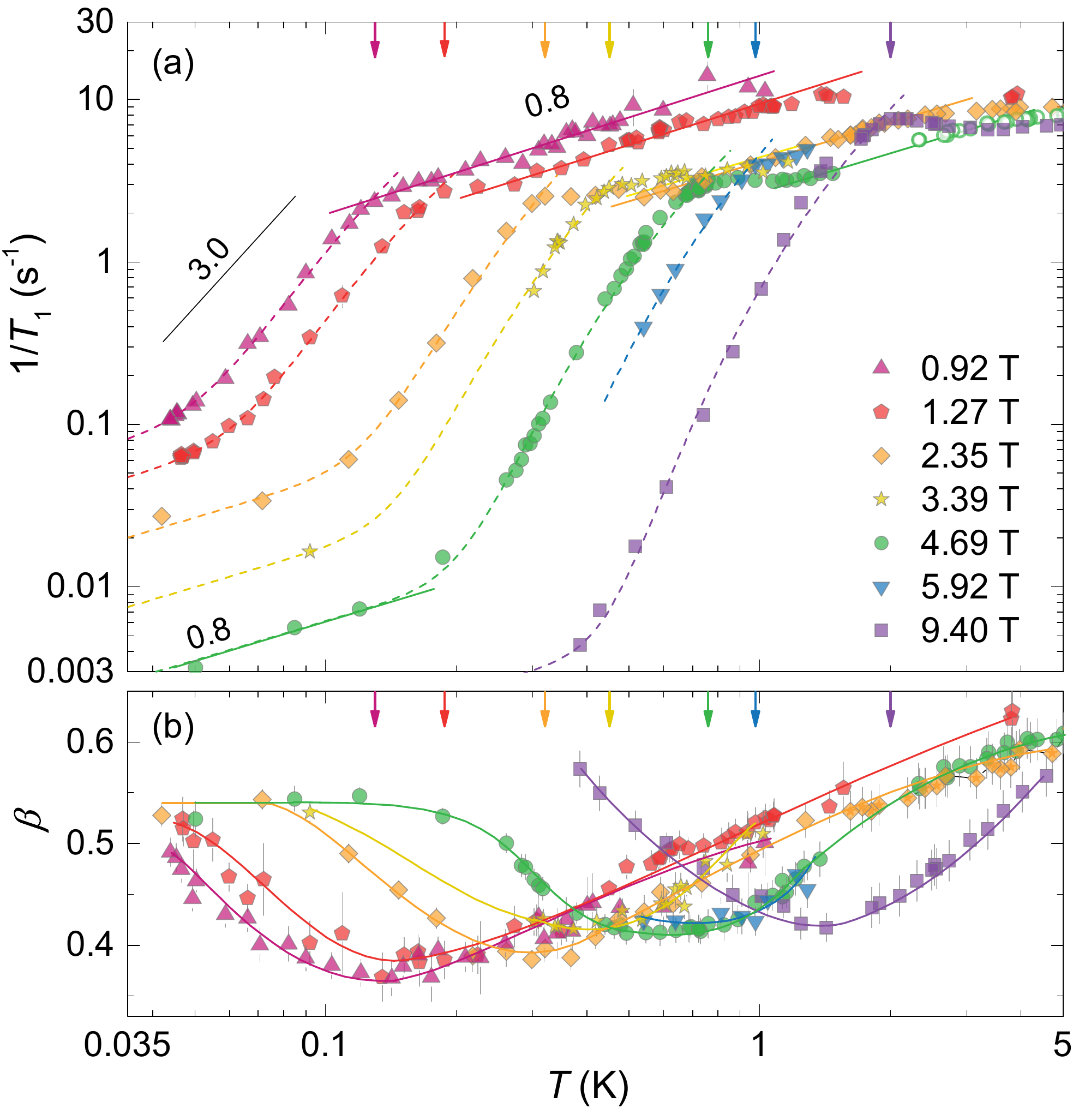}
\caption{
The temperature dependence of (a) the $^2$D NMR spin-lattice relaxation rate $1/T_1$ and (b) the stretching exponent $\beta$ from the magnetization recovery curves \cite{sup} in various applied fields.
In panel (a) the solid lines demonstrate power-law dependence (numbers correspond to powers), while
the dashed lines show the agreement with the gapped model of Eq.~(\ref{gap}).
We note that $T_{\rm c}$ roughly corresponds to a common relaxation rate $1/T_1=2.5$~s$^{-1}$ for all except the highest fields.
In panel (b) the solid lines are guides to the eye.
Arrows in all panels indicate the critical temperatures $T_{\rm c}$.
}
\label{fig2}
\end{figure}

In this study, we extend our $^2$D NMR $1/T_1$ data previously recorded in a field of 4.69 T between 2 and 300~K \cite{gomilsek2016instabilities}, down to 50~mK [Fig.~\ref{fig1}(a)].
As already established, the power-law decrease of $1/T_1$ below 200~K reveals  quantum critical behavior, while the decrease below the broad maximum around 5~K corresponds to the crossover within the SL state of \bro~\cite{gomilsek2016instabilities}.
Unexpectedly, by lowering the temperature to the millikelvin range we find a pronounced anomaly in the form of a sharp kink in $1/T_1$ at $T_{\rm c}=0.76$~K [Fig.~\ref{fig1}(a)], which is obviously not related to the smooth crossover within the SL state.
Below $T_{\rm c}$, $1/T_1$ quickly drops by several orders of magnitude [Fig.~\ref{fig2}(a)].
This is usually a sign of a phase transition and was encountered before in herbertsmithite \cite{jeong2011field} as well as in organic \cite{shimizu2006emergence, itou2010instability} and inorganic \cite{khuntia2016spin} triangular antiferromagnets.
However, $T_{\rm c}$ does not seem to correspond to a standard magnetic transition, where a divergence of $1/T_1$ is usually observed. 

The first essential question that arises is whether the magnetic state of \bro~below $T_{\rm c}$ remains a spin liquid. 
In order to address it we compare the NMR spectra recorded just above and well-below $T_{\rm c}$ [Fig.~\ref{fig1}(b)], as the emergence of frozen moments is generally reflected in a broadening of NMR lines.
This is indeed the case in $\kappa$-(BEDT-TTF)$_2$Cu$_2$(CN)$_3$ \cite{shimizu2006emergence} and in herbertsmithite in high magnetic fields \cite{jeong2011field}. 
On the other hand, the absence of any spectral broadening in EtMe$_3$Sb[Pd(dmit)$_2$]$_2$ was regarded as evidence of a non-trivial symmetry breaking and/or topological ordering within the SL state \cite{itou2010instability}, while no simultaneous anomaly was observed in any thermodynamic observable \cite{watanabe2012novel}.
Similarly to this case, the estimated magnetic contributions to the NMR line width \cite{sup} of 28.8(9) and 28.7(7)~mT at 0.26 and 0.90~K, respectively, demonstrate the absence of any magnetic broadening in \bro~upon crossing $T_{\rm c}=0.76$~K.
The uncertainty of the line widths and the hyperfine coupling constant between the $^2$D nuclei and kagome spins \cite{gomilsek2016instabilities}, $A^{\rm iso}=34$~mT/$\mu_{\rm B}$, set a very conservative upper bound of $0.05\mu_{\rm B}$ on the average frozen moment in the low-temperature state of \bro~at 4.69~T.
No spectral broadening is present even at 9.4~T \cite{sup}.
We thus arrive at the first important conclusion -- the dramatic change of the $1/T_1$ behavior in \bro~at $T_{\rm c}$ is {\it not due to bulk spin freezing}, but rather suggests a fundamental modification of the excitation spectrum of the SL state.

\begin{figure}[t]
\includegraphics[trim = 0mm 0mm 0mm 0mm, clip, width=1\linewidth]{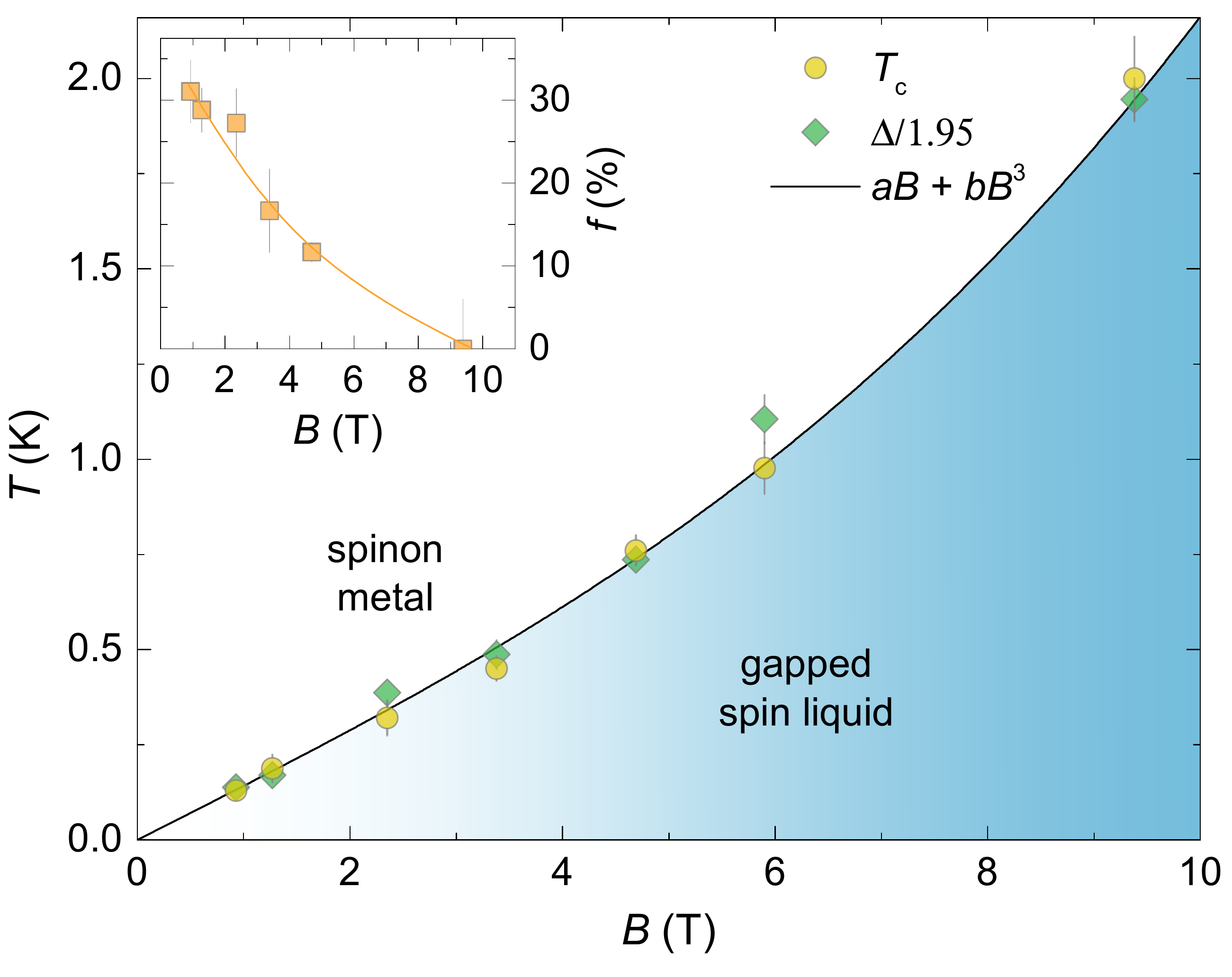}
\caption{The phase diagram of \bro~with the measured critical temperatures $T_{\rm c}$ and gaps $\Delta$ extracted from the low-$T$ $1/T_1$ NMR data. 
The line shows the $T_{\rm c}= aB+bB^3$ model fit of the phase boundary between two distinct spin liquid states.
The inset shows the field evolution of the fraction $f$ of the residual density of states at the Fermi level [Eq.~(\ref{gap})] below $T_{\rm c}$. The line in the inset is a guide to the eye.
}
\label{fig3}
\end{figure}

The next obvious questions are what triggers the instability at $T_c$ and whether it  is intrinsic to the kagome spins.
In order to address them, we performed additional NMR measurements in various magnetic fields between 0.92 and 9.4~T (Fig.~\ref{fig2}). 
We find \cite{sup} that $T_{\rm c}$ increases steadily with field.
It also roughly coincides with the temperature where the $1/T_1$ distribution in each field is the broadest -- the stretching exponent $\beta$, characterizing the distribution of relaxation times governing the magnetization recovery in the NMR experiment \cite{sup}, exhibits broad minima [Fig.~\ref{fig2}(b)].
Up to $B\sim 5$~T, $T_{\rm c}$ scales almost linearly with the applied field (with $T_{\rm c}\rightarrow 0$ when $B\rightarrow 0$), while at higher fields a cubic correction term is needed and the transition temperature obeys a phenomenological expression $T_{\rm c}=aB+bB^3$ (Fig.~\ref{fig3}).
This extra term could be due to the proximity of $T_{\rm c}$ to the crossover temperature regime of the SL state at higher fields, where the free-spinon density of states is changing \cite{gomilsek2016instabilities}. 
Alternatively, the crossover regime itself could be shifted to higher temperatures at higher fields as the additional Zeeman energy could stabilize spinon pinning, and the extra cubic term could be intrinsic to the instability.
In any case, the transition at $T_{\rm c}$ cannot be accounted for by the polarization of impurities in the applied field.
This conclusion is based on the fact that the slope of the $T_{\rm c}(B)$ curve is very small (it corresponds to an effective $g$-factor of 0.2), its shape at higher fields is convex, and the bulk magnetic susceptibility $\chi_{\rm b}$ in 4.69~T shows no significant anomaly  at $T_{\rm c}=0.76$~K [Fig.~\ref{fig1}(c)] despite a large impurity contribution at low temperatures \cite{li_gapless_2014}.
We thus reach the next important conclusion that the transition observed in \bro~at $T_{\rm c}$ reflects an {\it intrinsic instability within the SL phase induced by the applied magnetic field}.

Therefore, we turn to a more elaborate analysis of the NMR relaxation rates. Just above $T_{\rm c}$, the power law 
\begin{equation}
1/T_1=c T^\eta
\label{power}
\end{equation}
with $\eta=0.8$ fits the experiment [Fig.~\ref{fig2}(a)], except for the highest field.
This power is very similar to $\eta=0.73(5)$ found in herbertsmithite above 1~K \cite{olariu200817,jeong2011field}.
On the other hand, this dependence is also rather close to the Korringa relation $1/T_1\propto T$, which holds for free fermions \cite{abragam1961principles,yoshida2009phase}. 
In the picture of gapless $U(1)$ spinons with a Fermi surface, deviations towards $\eta<1$ are expected \cite{itou2010instability} due to the coupling of spinons with emergent gauge field fluctuations \cite{metlitski2015cooper}.
The $1/T_1$ data sets above $T_{\rm c}$ thus agree with the low-temperature zero-field SL state of \bro~being a gapless SL with a Fermi surface  \cite{gomilsek2016muSR}.
Moreover, the field-dependent parameter $c$, which is in such a SL proportional to the squared spinon density of states at the Fermi level (like in an ordinary metal \cite{abragam1961principles}), decreases with increasing field.
This is in line with the recently developed picture of spinon metals, where the spectral weight in the dynamic spin structure factor is progressively shifted away from zero energy by an increasing applied field \cite{li2017detecting}.
 
The significantly altered temperature dependence of $1/T_1$ just below $T_{\rm c}$ suggests an abrupt change of spin excitations within the spin liquid and, consequently, a large change of the spinon Fermi surface. 
However, at the lowest temperatures the sub-linear dependence of $1/T_1$ ($\eta \simeq 0.8$) is recovered [Fig.~\ref{fig2}(a)].
The field-dependence of this relaxation is incompatible with an impurity scenario \cite{sup}.
Moreover, the power $\eta$ is found to be the same as above $T_{\rm c}$.
Therefore, we propose that this relaxation at the lowest temperatures is due to a residual density of states of gapless spinons at the Fermi surface.
This is further supported by the decreasing relative width of the $T_1$ distribution at the lowest temperatures -- the stretching exponent increases back to $\beta\sim0.5$ [Fig.~\ref{fig2}(b)], the same value that characterizes the data above $T_{\rm c}$ \cite{sup}.

The coexistence of two relaxation mechanisms below $T_{\rm c}$ could, in principle, be due to an inhomogeneous phase (e.g., induced by disorder, structural imperfections, etc.), where regions of fully gapped and gapless spinons would coexist in real space \cite{galitski2007spin}.
However, this is not the case, as the magnetization recovery curves in the $T_1$ experiment should then exhibit a characteristic two-step shape at $T<T_{\rm c}$ where the relaxation times of the two phases differ significantly.
On the contrary, the experimental curves are smooth at all temperatures \cite{sup}.  
Therefore, a plausible explanation for the presence of two relaxation mechanisms below $T_{\rm c}$ is that the gap is opened only in certain regions of the momentum space \cite{galitski2007spin,lee2007amperean}.
Then, an effective two-channel relaxation can appear even in a spatially homogeneous state, because $1/T_1$ averages the dynamic spin structure factor over the entire momentum space \cite{moriya1956nuclear}, as $(T_1T)^{-1}\propto \sum_{\bf q} A_{\bf q}^2\chi''({\bf q}, \omega_{\rm NMR})/\omega_{\rm NMR}$, with $\chi''({\bf q}, \omega_{\rm NMR})$ being the imaginary part of the magnetic susceptibility at the wave vector ${\bf q}$ and NMR frequency $\omega_{\rm NMR}$.

The $1/T_1$ data at $T<T_{\rm c}$ can be accounted for by an extended thermal-activation model \cite{foot}
\begin{equation}
\frac{1}{T_1}= d\, T{\rm e}^{-\Delta/T}+f^2\cdot c T^{\eta},
\label{gap}
\end{equation}
%
where $d$ is a field-dependent parameter related to the relaxation mechanism induced by the opening of the gap $\Delta$, $c$ is determined from the data above $T_{\rm c}$ using Eq.~(\ref{power}) and $f < 1$ denotes the fraction of the residual density of states at the Fermi level with respect to the full density of states above $T_{\rm c}$ [Fig.~\ref{fig2}(a)]. 
At the lowest field of 0.93~T, the residual gapless-spinon density of states at $T<T_{\rm c}$ is $f\sim30$\% of the density of states above $T_{\rm c}$.
With increasing field $f$ decreases, reaching zero around 10~T (inset in Fig.~\ref{fig3}).
An important conclusion is thus that the spinon instability at $T_{\rm c}$ affects the {\it majority} of spinons near the Fermi surface.
The extracted excitation gap scales linearly with the transition temperature (Fig.~\ref{fig3}), $2\Delta/T_{\rm c} = 3.9(1)$, similarly as found in herbertsmithite \cite{jeong2011field} and close to the characteristic scaling $2\Delta/T_{\rm c} = 3.5$ of the BCS state.


Having established that the instability observed in \bro~at $T_{\rm c}$ is intrinsic and pertinent to the SL state, the important question of the underlying instability mechanism arises.
Two scenarios based on the current theoretical understanding of the spinon-metal phases in two dimensions could, in principle, be possible. 
An instability that gaps out a part or the whole Fermi surface is expected in the vicinity of the Mott transition where charge fluctuations are strong \cite{zhou2013spin, motrunich2006orbital}. 
Since in \bro~$U/t \sim 60$ \cite{sup} ($t$ and $U$ are the Hubbard hopping and the Coulomb repulsion, respectively), while the Mott metal-insulator transition occurs already at $U/t \gtrsim 10$ \cite{ohashi2006mott}, the system is positioned deep in the insulating phase. 
Therefore, we propose the second possible scenario, which is based on {\it a spinon-pairing instability} \cite{galitski2007spin,lee2007amperean,grover2010weak,metlitski2015cooper}.

The $U(1)$ SL's with spinon Fermi surfaces are found as root states of many $\mathbb{Z}_2$ fermionic SL's, when the spinon pairing amplitudes in the $\mathbb{Z}_2$ states are turned off \cite{chern2017fermionic}.
If spinon pairing that arises from the gauge-field mediated attractive interactions between spinons is considered, such states are susceptible to various pairing instabilities \cite{lee2007amperean, galitski2007spin, metlitski2015cooper}.
A paired-spinon state breaks the $U(1)$ gauge symmetry down to $\mathbb{Z}_2$ and opens an energy gap below the pairing temperature $T_{\rm c}$ \cite{galitski2007spin,grover2010weak}. 
The gap can either be full or partial, the latter corresponding to pairing in a non-symmetric channel \cite{lee2007amperean, metlitski2015cooper, grover2010weak}.
There are various possible types of the attractive interaction between spinons that lead to pairing instabilities \cite{galitski2007spin,lee2007amperean,metlitski2015cooper}.
If the spinon pairing were of the singlet BCS type, the paired state should be destroyed above a critical external magnetic field \cite{metlitski2015cooper}.
However, this is not the case for spin-triplet BCS pairing \cite{galitski2007spin} or a more exotic Amperean-type pairing with a spatially modulated amplitude \cite{lee2007amperean}.
The latter scenarios are therefore more likely to be realized in \bro, as our results suggest that the applied field stabilizes the gapped SL state and acts against thermal fluctuations which tend to prefer the gapless SL (Fig.~\ref{fig3}).

\bro~behaves very similarly to the organic triangular-lattice compound EtMe$_3$Sb[Pd(dmit)$_2$]$_2$, as both seem to exhibit a gapless quantum-critical spinon-metal ground state in zero magnetic field \cite{watanabe2012novel}
that undergoes an unconventional field-induced instability at low temperatures \cite{itou2010instability}.
In both systems, a gap develops in the excitation spectrum below a field-dependent critical temperature, even though the structure of the gap (full or nodal) may be different due to the different lattices of the two compounds. 
This instability, which is obviously not a regular thermodynamic transition into a frozen spin state, but rather suggests that the magnetic field strongly affects the spinon excitations, thus seems to be a {\it general characteristic} of SL's with spinon Fermi surfaces.
Therefore, it would be interesting to check for its presence in other physically different spinon-metal ground-state candidates, one of them possibly being the rare-earth based triangular-lattice antiferromagnet YbMgGaO$_4$ \cite{shen2016evidence, li2015gapless, paddison2016continuous, li2017nearest}.

Another interesting aspect of the discovered field-induced instability in \bro~is that an infinitesimal field apparently suffices to destabilize the zero-field spinon Fermi-surface SL ground state.
This is compatible with its gapless nature and stands in contrast to gapped SL's where a finite critical field is generally expected at zero temperature \cite{watanabe2012novel}.
In this respect, the SL with a spinon Fermi surface responds to the applied field in a manner similar to that theoretically predicted for some other gapless SL's; e.g., in an infinitesimal magnetic field, the gapless Dirac $U(1)$ SL on the kagome lattice should become unstable towards spontaneous spin ordering with gapped excitations \cite{ran2009spontaneous} and the gapless Majorana fermions of the Kitaev model on the honeycomb lattice become gapped \cite{kitaev2006anyons, jansa2017observation}.  

Beyond the intricate physics of spin liquids, our finding of an intrinsic field-induced instability of the gapless Fermi-surface SL realized in \bro~should turn out to be relevant in a broader context of the Fermi-surface instabilities. 
It is important to bear in mind that many aspects of pair condensation in such SL's in Mott insulators are {\it universal}, as they are shared by electron pairing in superconductors \cite{galitski2007spin}, electron pairing in metals mediated by order-parameter fluctuations near quantum critical points, and composite-fermion pairing in quantum Hall fluids \cite{metlitski2015cooper}.
It is even possible that in some systems the same fermionic pairing occurs in both SL and superconducting phases \cite{powell2011quantum}. 

\acknowledgments{The financial support of the Slovenian Research Agency under the project BI-FR/15-16-PROTEUS-004 and the program No.~P1-0125 is acknowledged. M.~G.~thanks the Slovene Human Resources Development and Scholarship Fund for financial support (contract No. 11012-8/2015).
This work was also supported by the French Agence Nationale de la
Recherche under ``SPINLIQ" Grant No. ANR-12-BS04-0021 and by Universit\'e Paris-Sud Grant MRM PMP.
Q.~M.~Z. was supported by the NSF of China and the Ministry of Science and Technology of China (973 projects: 2016YFA0300504).
%
%
%
%

\begin{widetext}
\vspace{19cm}
\begin{center}
{\large {\bf Supplementary information:\\
Field-Induced Instability of a Gapless Spin Liquid with a Spinon Fermi Surface}}\\
\vspace{0.5cm}
M. Gomil\v sek,$^{1}$ M. Klanj\v sek,$^{1}$ R. \v Zitko,$^{1}$ M. Pregelj,$^1$ F. Bert,$^{2}$ P. Mendels,$^{2}$ Y. Li,$^{3}$ Q. M. Zhang,$^{3, 4}$ and A. Zorko$^{1,*}$
\vspace{0.3cm}

{\it
$^1$Jo\v{z}ef Stefan Institute, Jamova c.~39, SI-1000 Ljubljana, Slovenia\\
\vspace{0.1cm}
$^2$Laboratoire de Physique des Solides, UMR CNRS 8502, Universit\'e Paris Sud, 91405 Orsay, France\\
\vspace{0.1cm}
$^3$Department of Physics, Renmin University of China, Beijing 100872, People's Republic of China\\
\vspace{0.1cm}
$^4$Department of Physics and Astronomy, Shanghai Jiao Tong University, Shanghai 200240 and Collaborative Innovation Center of Advanced Microstructures, Nanjing 210093, P. R. China\\
}

\end{center}
\end{widetext}

\section{Nuclear Magnetic Resonance}\label{appA}
\subsection{NMR Experiment}\label{appA1}
The $^{2}$D ($I=1$) nuclear magnetic resonance (NMR) measurements were performed on the same powder sample as in our previous NMR study that was limited to temperatures above 2~K and to a single applied magnetic field of 4.69~T \cite{gomilsek2016instabilities}.
To reach temperatures below 1.5~K, a dilution-refrigerator setup was used, together with a sweepable superconducting magnet capable of reaching 14~T.
The $^2$D NMR spectra were recorded by sweeping the magnetic field.
The standard solid-echo pulse sequence $\pi/2-\tau-\pi/2$ was used, with a typical pulse length and pulse separation of 10~$\mu$s and 120~$\mu$s, respectively.
The spin-lattice relaxation time $T_1$ was measured after the $\pi/2-t-\pi/2$ sequence with the solid-echo detection sequence.

\subsection{Streched-Exponential Relaxation}\label{appA2}
\begin{figure}[h]
\includegraphics[trim = 0mm 0mm 0mm 0mm, clip, width=1\linewidth]{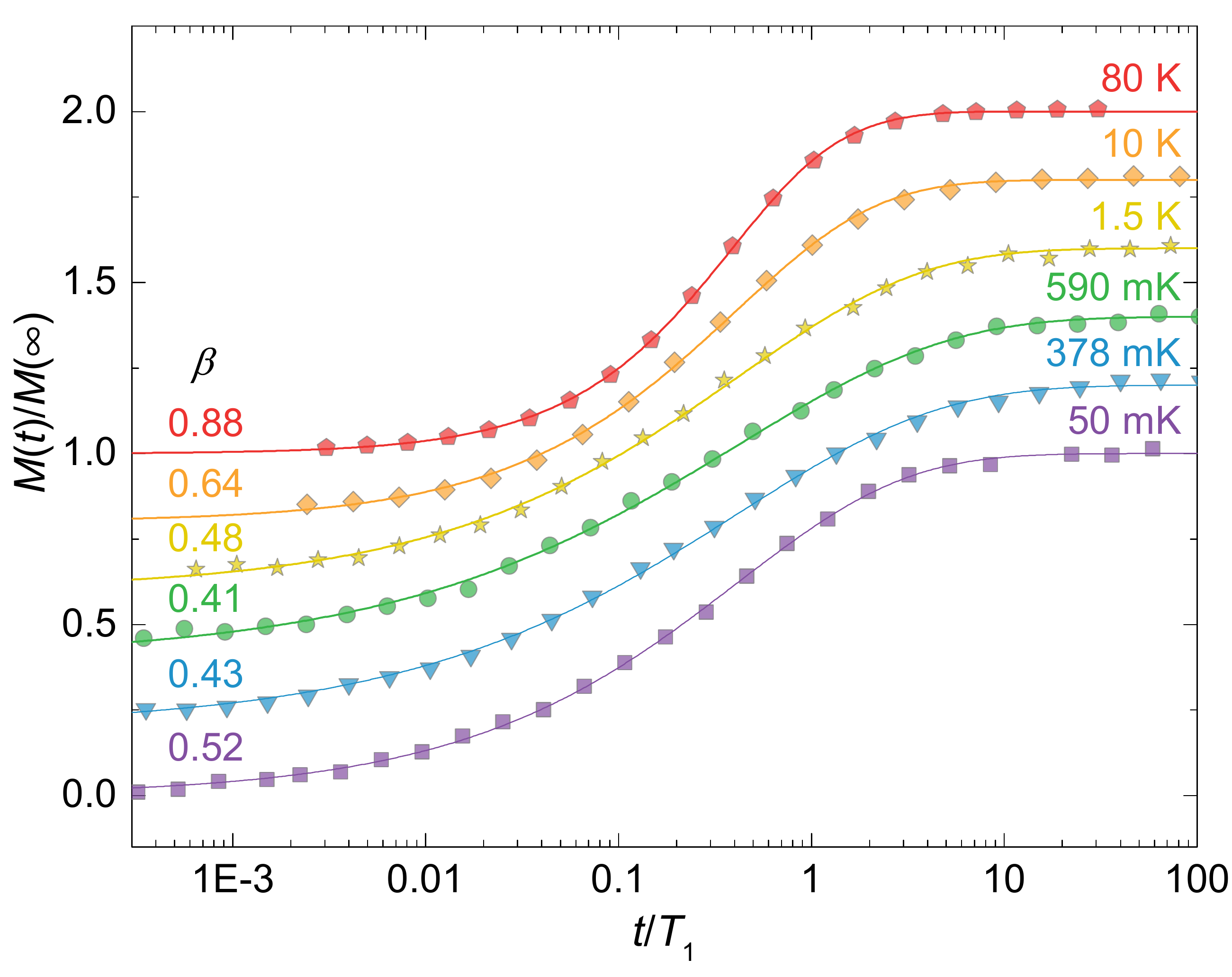}
\caption{The normalized magnetization recovery curves at several selected temperatures (symbols) and the corresponding fits  with Eq.~(\ref{stretch}) (solid lines; the stretching exponent values $\beta$ are also given). The curves are shifted vertically for clarity.
}
\label{figS1}
\end{figure}
The magnetic-relaxation model for spin-1 nuclei \citep{suter1998mixed}, 
\begin{equation}
\label{stretch}
M(t)=M_0\left[1-\left(1-s\right)\left(\frac{1}{4} {\rm e}^{-(t/T_1)^\beta} + \frac{3}{4} {\rm e}^{-(3t/T_1)^\beta}  \right) \right],
\end{equation}
was fitted to the magnetization recovery curves in the spin-lattice relaxation experiments.
Here $M_0$ is the saturation magnetization, $s\neq 0$ accounts for imperfect saturation of broad NMR lines after the first $\pi/2$ pulse and $\beta$ denotes a stretched exponential relaxation.
Eq.~(\ref{stretch}) perfectly fits the experimental curves in the whole temperature range between 300~K and 50~mK (Fig.~\ref{figS1}).
In particular at $T<T_{\rm c}$, where two relaxation mechanisms are present (see main text), we do not observe a two-step magnetization recovery curve at any temperature (even when the relaxation times of the two mechanisms are different by two orders of magnitude), which would otherwise imply a spatially inhomogeneous response.
\begin{figure}[b]
\includegraphics[trim = 0mm 0mm 0mm 0mm, clip, width=1\linewidth]{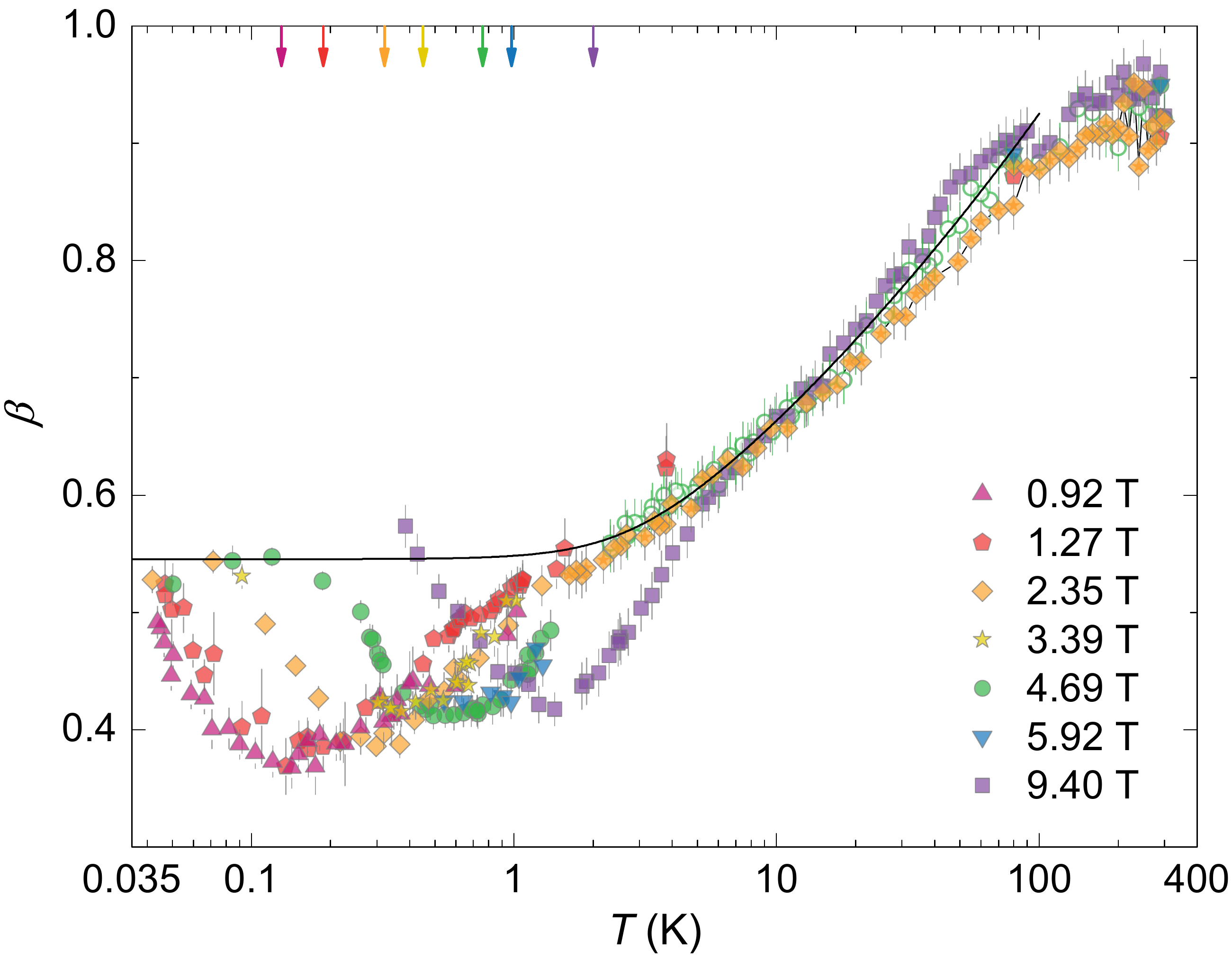}
\caption{The temperature dependence of the stretching exponent $\beta$ from fitting the model (\ref{stretch}) to the experimental magnetization recovery curves (Fig.~\ref{figS1}). 
The high-$T$ data in 4.69~T (open symbols) is reproduced from Ref.~\onlinecite{gomilsek2016instabilities}.
The thick line is a guide to the eye and shows the trend of the field-independent $\beta(T)$, which is violated around $T_{\rm c}$ (denoted by arrows) and again recovered at low temperatures.
}
\label{figS2}
\end{figure}

The stretching exponent $\beta$ was found to be strongly $T$-dependent.
However, this parameter does not depend on the applied field (Fig.~\ref{figS2}), except close to $T_{\rm c}$, where $\beta$ exhibits a minimum at each applied field.
It monotonically decreases from the room-temperature value of 0.93(2), reaches its low-$T$ value of 0.54(2) just above $T_{\rm c}$ and returns to this value at the lowest temperatures.

\subsection{Determination of $T_{\rm c}$}\label{appC1}
The critical temperature $T_{\rm c}$ for the transition between the gapless spinon Fermi-surface SL and the gapped SL is indicated by a kink in the $1/T_1$ temperature dependence [Fig.~2(a) in the main text] above which the gapped model of Eq.~(2) in the main text is no longer valid. To obtain an objective estimate of $T_{\rm c}$ we fitted the low-temperature $1/T_1$ data with this gapped model up to different cut-off temperatures and calculated the chi-squared goodness of fit statistics (Fig.~\ref{figS5}). We then defined $T_{\rm c}$ as the maximum temperature where Pearson's chi-squared test still indicated a good fit at a 95\% confidence level.
\begin{figure}[h]
\includegraphics[trim = 0mm 0mm 0mm 0mm, clip, width=1\linewidth]{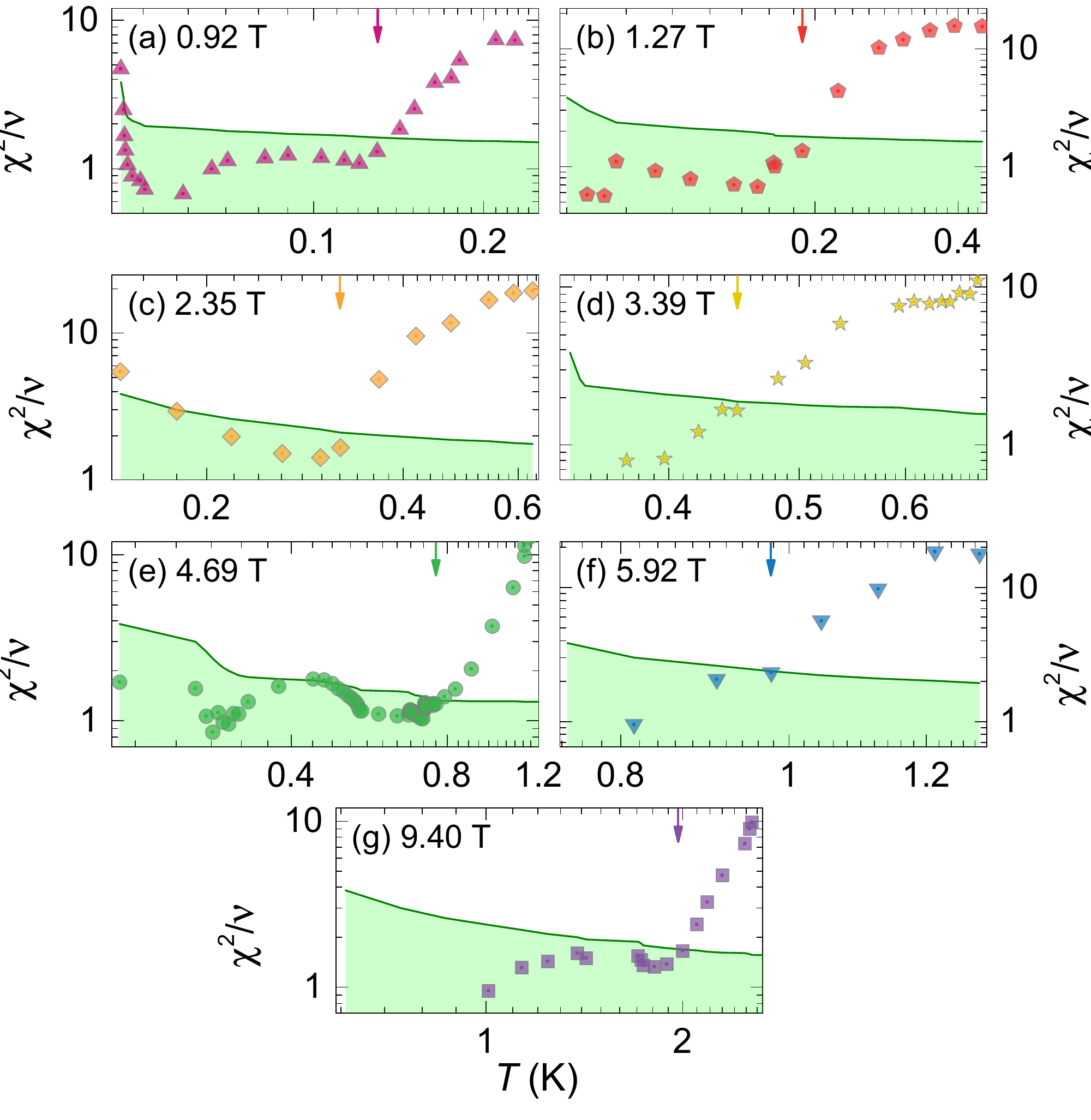}
\caption{Reduced chi-squared $\chi^2/\nu$ (symbols), where $\nu$ is the number of degrees of freedom, versus the cutoff temperature $T$ for fits of low-temperature $1/T_1$ data using the gapped model of Eq.~(2) in the main text. The green shaded areas indicate $\chi^2/\nu$ values for a good fit at a 95\% confidence level via Pearson's chi-squared test. Arrows indicate the critical temperatures $T_{\rm c}$ up to which the test still indicates a good fit.}
\label{figS5}
\end{figure}
%

%
%
%
\begin{figure}[b]
\includegraphics[trim = 0mm 0mm 0mm 0mm, clip, width=1\linewidth]{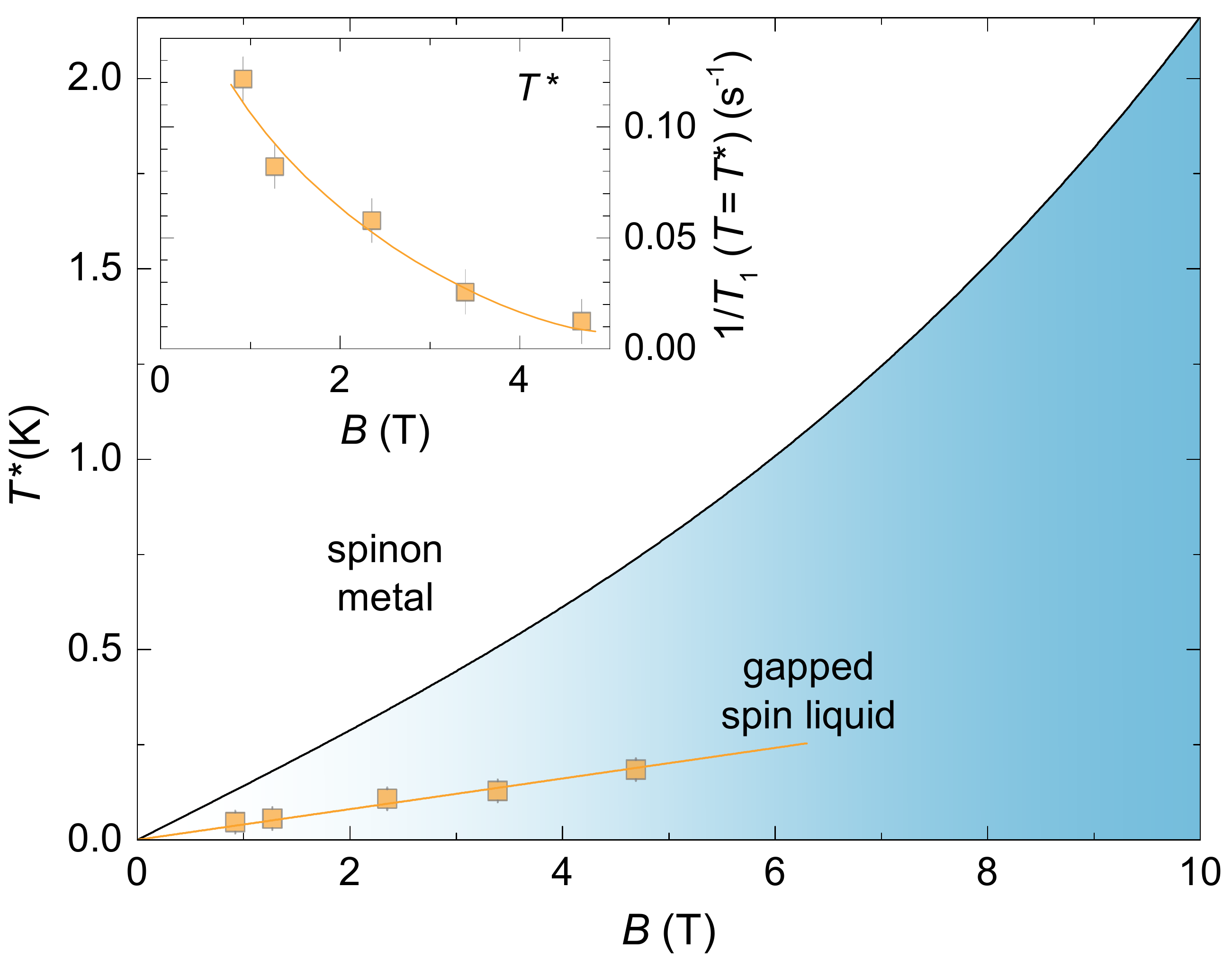}
\caption{The field dependence of the onset temperature $T^*$ of the low-temperature relaxation mechanism. 
Inset shows the decrease of the NMR relaxation rate $1/T_1$ at $T^*$ with increasing field. 
}
\label{figS3}
\end{figure}

\subsection{Low-$T$ Relaxation}\label{appA3}
The NMR relaxation rate at the lowest temperatures decays much slower with decreasing temperature than just below the transition temperature $T_{\rm c}$ [Fig.~2(a) in the main document]. 
This is a fingerprint of a second relaxation mechanism that takes over at the lowest temperatures, where the exponentially decaying relaxation mechanism due to an excitation gap below $T_{\rm c}$ becomes weaker.
The crossover temperature $T^*<T_{\rm c}$ is defined as the temperature, where both relaxation mechanisms match (Fig~\ref{figS4}).
The emergence of the low-$T$ relaxation channel slowly varying with temperature, which is regularly observed when the main relaxation channel follows an activation-type behavior and the relaxation rates severely drop at low temperatures, is usually ascribed to impurities.
However, we argue that impurities are not responsible for the low-$T$ sub-linear NMR relaxation in \bro.
Any impurities significantly coupled to the gapped spinons of the field-induced SL state would exhibit an activation-type behavior.
On the other hand, the relaxation rate of free impurities should follow some functional dependence $1/T_1^{\rm imp}=f(B/T)$ \cite{kermarrec2011spin}.
Hence, at $T^*$ the measured relaxation rate due to both equally contributing relaxation channels would be $1/T_1(T^*,B)=2 f(B/T^*)$.
Since $T^*(B)/B$ is constant (Fig.~\ref{figS3}), $1/T_1(T^*,B)$ should be $B$-independent.
This contradicts the experiment, where an increasing magnetic field strongly suppresses $1/T_1(T^*,B)$ (inset in Fig.~\ref{figS3}).
Therefore, we conclude that the sub-linear relaxation term that is found at the lowest temperatures is due to intrinsic relaxation originating from kagome-lattice excitations.
This is supported by the fact that the same power $\eta=0.8$ is found for this relaxation as for the relaxation above $T_{\rm c}$.

\subsection{Power-law Relaxation below $T_{\rm c}$}\label{appA4}
The temperature dependence of $1/T_1$ below the critical temperature $T_c$, which is in the main document fitted with the extended thermal-activation model, can be, alternatively, modeled with a double-power-law model (see Fig.~\ref{figS4})
\begin{equation}
\frac{1}{T_1}= d'\, T^{\eta'}+f^2\cdot c T^{\eta},
\label{2power}
\end{equation}
where the exponentially depending term $T{\rm e}^{-\Delta/T}$ (see main text) is replaced by the power-law term $T^{\eta'}$.
Here, $\eta=0.8$ is found again, while the $\eta'$ parameter exhibits a pronounced field dependence (inset in Fig.~\ref{figS4}).
In the limit $B\rightarrow 0$ we find $\eta'\rightarrow 3.2(2)$.
\begin{figure}[h]
\includegraphics[trim = 0mm 0mm 0mm 0mm, clip, width=1\linewidth]{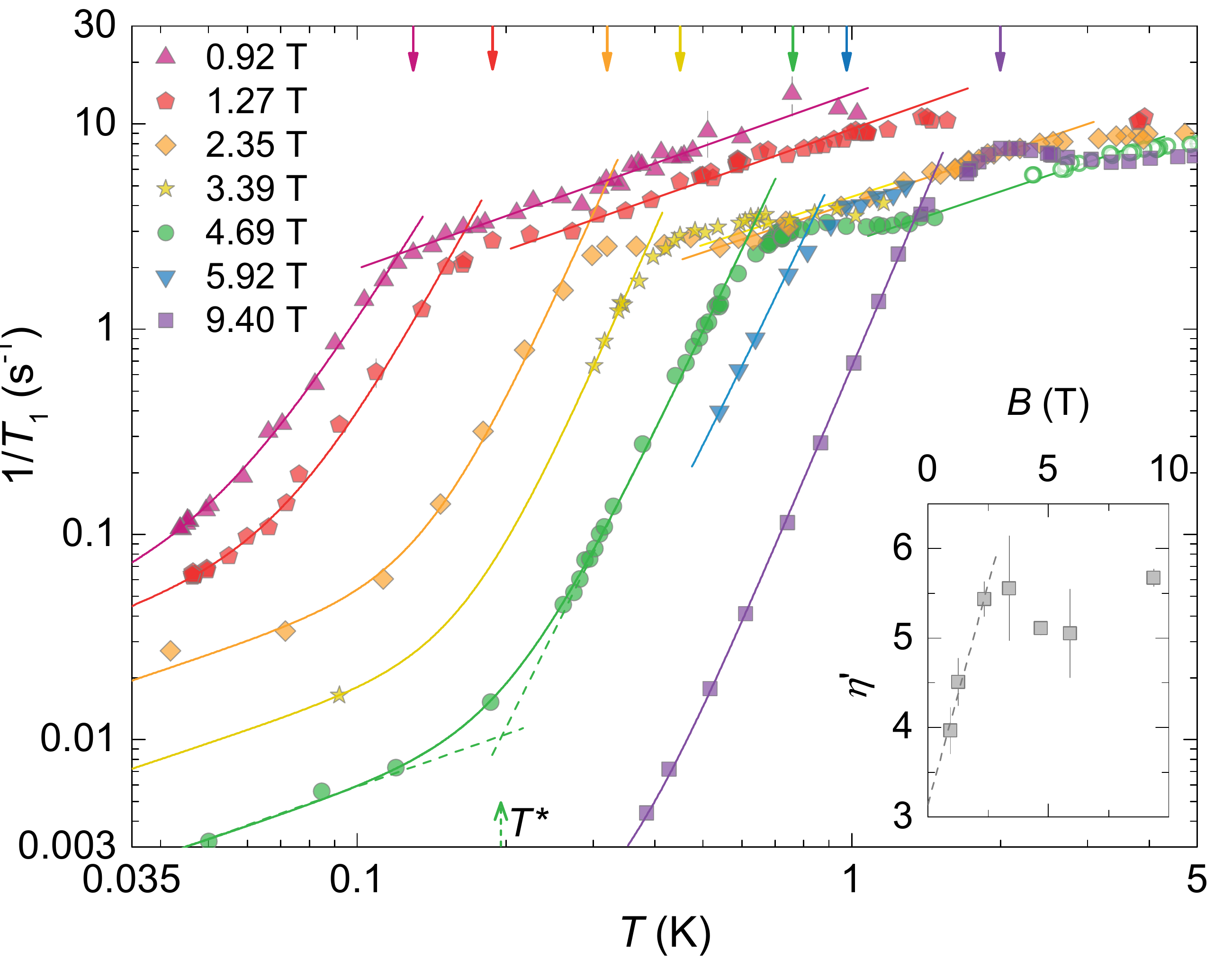}
\caption{The temperature dependence of the spin-lattice relaxation rate $1/T_1$ (symbols) in various fields.
Below $T_{\rm c}$, the solid lines are fits with the double-power-law model of Eq.~(\ref{2power}) denoted by solid arrows.
Above $T_{\rm c}$ the lines demonstrate the $1/T_1\propto T^{0.8}$ dependence.
The dashed lines are extrapolations of both low-$T$ relaxation regimes that are used to define $T^*$. 
The inset shows the field-dependent power $\eta'$ of the fast relaxing component in Eq.~(\ref{2power}), a dashed line is used for the extrapolation to B=0.
}
\label{figS4}
\end{figure}

\subsection{NMR Broadening}\label{appA5}
At 4.69~T the magnetic contribution to the NMR line width was obtained from the second central moment of the NMR spectra, $m_2=\overline{\nu^2}-\overline{\nu}^2$, as $\sigma_{\rm m}=\sqrt{m_2-\sigma_\infty^2}$, where $\sigma_\infty^2\approx m_2(300$~K$)$ is related to quadrupolar broadening \cite{gomilsek2016instabilities}. The spectra were windowed to the $0.015\sim0.985$ quantile range to reduce the influence of statistical noise in the line tails. The experimental uncertainty in $\sigma_{\rm m}$ cited in the main text is ultimately dominated by the systematic uncertainty in the choice of the windowing range.

To check for any spectral broadening upon crossing the critical temperature $T_{\rm c}=2.0$~K at 9.4~T we measured spectra down to 1.73~K, where the NMR line width $m_2^{1/2}$ approaches saturation (Fig.~\ref{figS6}). The line width does not seem to be affected at $T_{\rm c}$. However, as 1.73~K is still rather close to $T_{\rm c}$ we also extracted the saturation magnetizations $M_0$ from $1/T_1$ measurements using Eq.~(\ref{stretch}), in the temperature range between 300 K and 385~mK. Below 4~K the saturation magnetization follows a simple Curie law $M_0 \propto 1/T$ down to the lowest temperature (Fig.~\ref{figS6}). This is expected for the total NMR intensity due to Boltzmann statistics of equilibrium nuclear spin populations in an external magnetic field. It implies that the underlying NMR spectrum does not exhibit any noticeable broadening across the instability even at 9.4~T, as a changing NMR linewidth would result in a deviation from the $1/T$ temperature dependence for the saturation magnetization.
Such deviation towards higher values is observed above 4~K, where the spectra narrow notably with increasing temperature.
The diverging trend of the saturation magnetization from the $1/T$ dependence is altered above $\sim$100~K, where the magnetic broadening becomes masked by the quadrupolar broadening, completely changing the shape of the NMR spectrum (inset in Fig.~\ref{figS6}).
\begin{figure}[t]
\includegraphics[trim = 0mm 0mm 0mm 0mm, clip, width=1\linewidth]{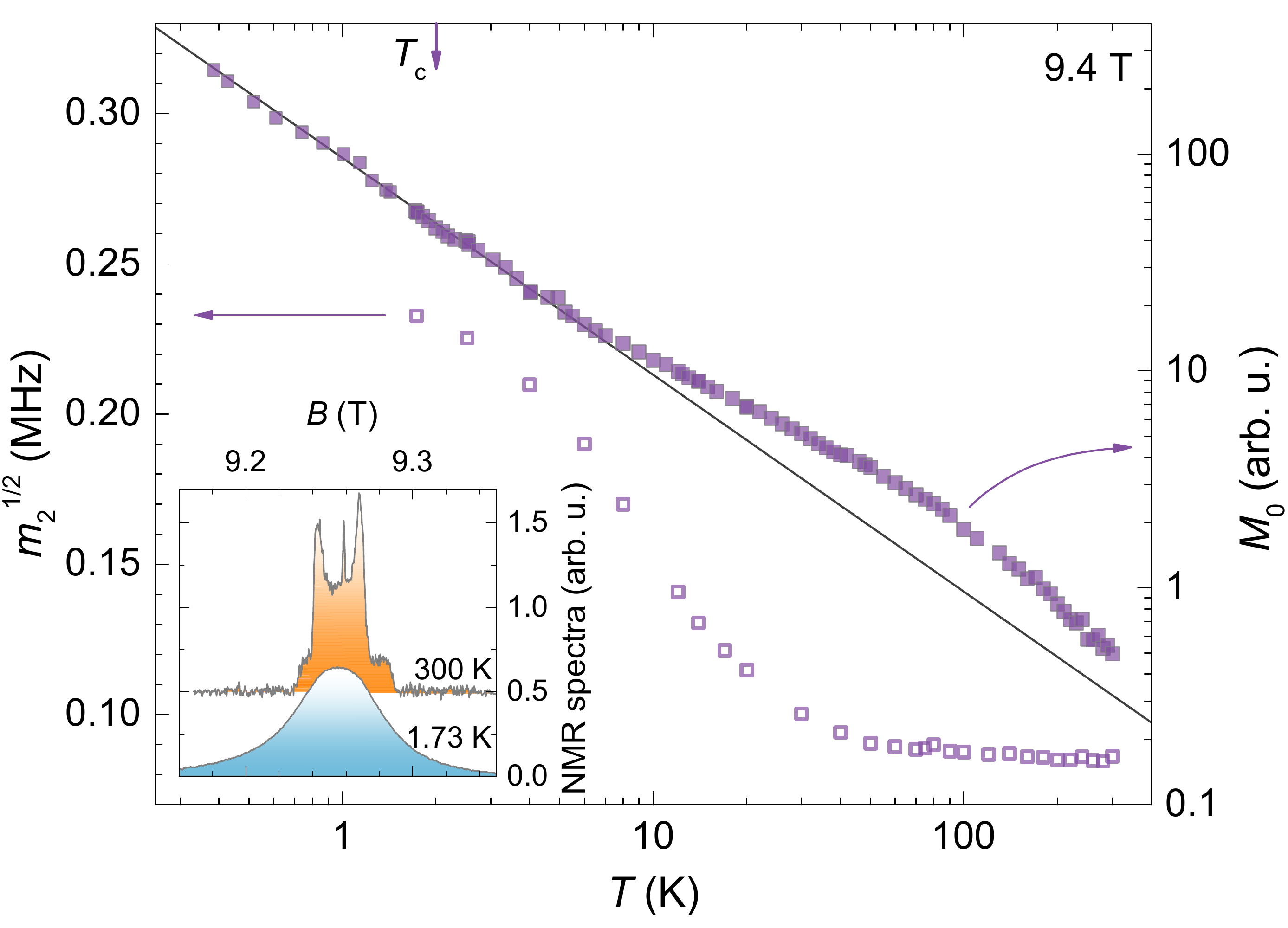}
\caption{The temperature dependence of the NMR line width ${m_2^{1/2}}$ (empty symbols) and the saturation magnetization $M_0$ from $1/T_1$ measurements at 9.4~T (solid symbols). The line indicates a Curie temperature dependence. The arrow indicates the critical temperature $T_{\rm c}$. The inset shows normalized NMR spectra at the highest and the lowest temperature, which are shifted vertically for clarity.}
\label{figS6}
\end{figure}

\section{Mott Transition}\label{appB}
Within the kagome lattice Hubbard model, the system undergoes a first-order Mott phase transition at the Hubbard repulsion $U/W\sim 1.4$, where $W=6t$ represent the bandwidth, $t$ is the hopping parameter and $U$ is the Coulomb repulsion \cite{Ohashi2006Mott}.
In \bro~the parameters $U$ and $t$ can be estimated \cite{ashcroft2010solid} from the measured bang gap $E_g=U-2zt=4.2$~eV ($z=4$) and the average nearest-neighbor exchange constant $k_{\rm B}J=4t^2/U=5.6$~meV \cite{li_gapless_2014}.
We find $U=4.9$~eV and $t=0.08$~eV, the former being close to the value $U=6$~eV determined from in-depth band-structure calculations of the structurally similar  herbertsmithite and kapellasite, ZnCu$_3$(OH)$_6$Cl$_2$ \cite{Jeschke2013First}.
The resulting ratio $U/W\sim 10$ means that \bro~is located deep in the insulating phase far away from the Mott transition. 

%

\end{document}